\documentclass[12pt]{article}
\begin{document}

\begin{center}
{\Large\bf{}A simple tensorial proof for the completely symmetric
property of the Bel-Robinson tensor}\\
(Dated on 9 June 2010, revised 9 July 2010, s0242010@gmail.com)
\end{center}

\begin{center}
Lau Loi So\\
Department of Physics, National Central University, Chung Li 320,
Taiwan
\end{center}

\begin{abstract}
The Bel-Robinson tensor $T_{\alpha\beta\mu\nu}$ was proposed in
1958.  The main application of this tensor is for describing
gravitational energy.  It is known that $T_{\alpha\beta\mu\nu}$
has many nice properties such as being completely symmetric. It is
easy to prove this property using spinors as shown in Penrose's
book.  The main purpose of the present paper is to verify that the
Bel-Robinson tensor is indeed completely symmetric using a basic
tensorial method.  After we have this result we learned that
Senovilla in 2000 has already used the similar idea to obtain the
same result.  However, keep using the tensorial method, we propose
another easier proof that $T_{\alpha\beta\mu\nu}$ is indeed
totally symmetric. Moreover, we also found that the well known
equation in vacuum,
$R_{\alpha\lambda\sigma\tau}R_{\beta}{}^{\lambda\sigma\tau}
\equiv\frac{1}{4}g_{\alpha\beta}R_{\rho\lambda\sigma\tau}
R^{\rho\lambda\sigma\tau}$, which can be proven by the same
tensorial method.
\end{abstract}

\section{Introduction}
The famous Bel-Robinson tensor
$T_{\alpha\beta\mu\nu}$~\cite{Bel-Robinson} was proposed in 1958.
Nowadays it is called a superenergy tensor~\cite{Senovilla}. It is
believed that the gravitational field energy is related to
$T_{\alpha\beta\mu\nu}$~\cite{Szabados} as it gives a positivity
energy density quasilocally.  The quasilocal idea (i.e., within a
closed 2-surface) is physical, which means that the gravitational
energy density is well defined at the quasilocal level
theoretically~\cite{Hawking,Penrose1982}. There were many papers
in the literature using $T_{\alpha\beta\mu\nu}$ to describe the
gravitational
energy~\cite{Garecki1973,Horowitz,Krishnasamy,SoCQG2009}. The
Bel-Robinson tensor has many nice properties such as being
completely symmetric, totally trace free and divergence free. It
also fulfills the dominant energy condition. There are different
ways to define $T_{\alpha\beta\mu\nu}$, one of the common
expressions is
\begin{eqnarray}
T_{\alpha\beta\mu\nu}&:=&R_{\alpha\lambda\mu\sigma}R_{\beta}{}^{\lambda}{}_{\nu}{}^{\sigma}
+\ast{}R_{\alpha\lambda\mu\sigma}\ast{}R_{\beta}{}^{\lambda}{}_{\nu}{}^{\sigma}\nonumber\\
&=&R_{\alpha\lambda\mu\sigma}R_{\beta}{}^{\lambda}{}_{\nu}{}^{\sigma}
+R_{\alpha\lambda\nu\sigma}R_{\beta}{}^{\lambda}{}_{\mu}{}^{\sigma}
-\frac{1}{2}g_{\alpha\beta}R_{\mu\lambda\sigma\tau}R_{\nu}{}^{\lambda\sigma\tau}\nonumber\\
&=&R_{\alpha\lambda\mu\sigma}R_{\beta}{}^{\lambda}{}_{\nu}{}^{\sigma}
+R_{\alpha\lambda\nu\sigma}R_{\beta}{}^{\lambda}{}_{\mu}{}^{\sigma}
-\frac{1}{8}g_{\alpha\beta}g_{\mu\nu}R_{\rho\lambda\sigma\tau}R^{\rho\lambda\sigma\tau},
\end{eqnarray}
where $\ast{}R_{\alpha\lambda\mu\sigma}
=\frac{1}{2}\epsilon_{\alpha\lambda\xi\kappa}R^{\xi\kappa}{}_{\mu\sigma}$
is the dual of the Riemann curvature tensor and we have made use
the vacuum relation 
$R_{\alpha\lambda\sigma\tau}R_{\beta}{}^{\lambda\sigma\tau}
=\frac{1}{4}g_{\alpha\beta}R_{\rho\lambda\sigma\tau}R^{\rho\lambda\sigma\tau}$.
It is transparent to see some of the symmetry properties
\begin{equation}
T_{\alpha\beta\mu\nu}=T_{(\alpha\beta)(\mu\nu)}=T_{(\mu\nu)(\alpha\beta)},\label{10cJune2010}
\end{equation}
but the totally symmetric property is not obvious.  Thus Penrose
and Rindler~\cite{Penrose} say ``The symmetry properties of
$T_{\alpha\beta\mu\nu}$ are by no means apparent from the tensor
formula, but they follow directly from the spinor expression ...",
p241.  We already know that this tensor should be completely
symmetric because it is an analog of the symmetric trace-free
divergence-free tensor, the energy-momentum tensor for the
electromagnetic field,
\begin{equation}
T_{\alpha\beta}:=\frac{1}{2}(F_{\alpha\lambda}F_{\beta}{}^{\lambda}
+\ast{}F_{\alpha\lambda}\ast{}F_{\beta}{}^{\lambda})=
F_{\alpha\lambda}F_{\beta}{}^{\lambda}
-\frac{1}{4}g_{\alpha\beta}F_{\rho\lambda}F^{\rho\lambda},
\end{equation}
where $\ast{}F_{\alpha\beta}$ is the dual 2-form of
electromagnetic field strength tensor $F_{\alpha\beta}$.  As
$T_{\alpha\beta}$ possesses the dominant energy condition, then
$T_{\alpha\beta\mu\nu}$ should also.

It may need to be emphasized that the completely symmetric
property of $T_{\alpha\beta\mu\nu}$ is important
\cite{Senovilla,BergqvistCQG2004}. Penrose \cite{Penrose} used
spinors to verify that $T_{\alpha\beta\mu\nu}$ is really totally
symmetric long ago.  However, could it be possible to use the
traditional tensorial method to proof this nice property? This may
help someone who is not familiar with spinor techniques especially
for the beginner to study the general relativity.  The answer is
yes.  The main purpose of the present paper is to verify that the
Bel-Robinson tensor is indeed completely symmetric using a basic
tensorial method.  After we have this result we learned that
Senovilla~\cite{Senovilla} (see Proposition 6.3) in 2000 has
already used the similar idea to obtain the same result.  However,
insist using the tensorial method and making use the formal dual
of left and right, we propose another easier proof that
$T_{\alpha\beta\mu\nu}$ is indeed completely symmetric. Moreover,
using this symmetric property, we also found that the vacuum
relation
$R_{\alpha\lambda\sigma\tau}R_{\beta}{}^{\lambda\sigma\tau}
\equiv\frac{1}{4}g_{\alpha\beta}R_{\rho\lambda\sigma\tau}
R^{\rho\lambda\sigma\tau}$ can be obtained using the same method.

\section{Technical background}
In order to prove the completely symmetric property of
$T_{\alpha\beta\mu\nu}$, we need the following relation which is
only valid in vacuum,
\begin{equation}
\epsilon_{\rho\lambda\xi\kappa}R^{\xi\kappa}{}_{\sigma\tau}
+\epsilon_{\rho\sigma\xi\kappa}R^{\xi\kappa}{}_{\tau\lambda}
+\epsilon_{\rho\tau\xi\kappa}R^{\xi\kappa}{}_{\lambda\sigma}\equiv0,\label{10aJune2010}
\end{equation}
where $\epsilon_{\rho\lambda\xi\kappa}$ is the totally
skew-symmetric Levi-Civita tensor.  This equation looks like
making a dual on the first Bianchi identity
\begin{equation}
\ast{}R_{\rho\lambda\sigma\tau}+\ast{}R_{\rho\sigma\tau\lambda}+\ast{}R_{\rho\tau\lambda\sigma}\equiv0.
\label{29aJune2010}
\end{equation}
However, it is not true in general but only true in vacuum. The
detailed verification will be demonstrated in the next paragraph.
After some simple algebra using (\ref{10aJune2010}), the
anti-symmetric property of $\epsilon_{\alpha\beta\mu\nu}$ and
$R_{\alpha\beta\mu\nu}$, we can obtain one more relation
\begin{equation}
\epsilon_{\rho\lambda\xi\kappa}R^{\xi\kappa}{}_{\sigma\tau}
\equiv\epsilon_{\sigma\tau\xi\kappa}R^{\xi\kappa}{}_{\rho\lambda}\quad\leftrightarrow\quad
\ast{}R_{\rho\lambda\sigma\tau}\equiv\ast{}R_{\sigma\tau\rho\lambda}.\label{30aJune2010}
\end{equation}
Once again, this looks like a property of the Riemann curvature
tensor, but it is only valid in vacuum.  In fact, this result is
well known, the left dual and right dual and they are equal in
vacuum.  Explicitly
\begin{equation}
\epsilon_{\rho\lambda\xi\kappa}R^{\xi\kappa}{}_{\sigma\tau}
\equiv{}R_{\rho\lambda}{}^{\xi\kappa}
\epsilon_{\xi\kappa\sigma\tau}\quad\leftrightarrow\quad
\ast{}R_{\rho\lambda\sigma\tau}\equiv{}{R\ast}_{\rho\lambda\sigma\tau}.\label{30bJune2010}
\end{equation}

Here we verify the relation (\ref{10aJune2010}) using differential
forms.  Define the dual of the the curvature 2-form
$R^{\alpha\beta}$ as follows
\begin{equation}
(\ast{}R)_{\mu\nu}:=\frac{1}{2}\epsilon_{\mu\nu\xi\kappa}R^{\xi\kappa}
=\frac{1}{4}\epsilon_{\mu\nu\xi\kappa}R^{\xi\kappa}{}_{\lambda\sigma}\theta^{\lambda}\wedge\theta^{\sigma}.
\end{equation}
Consider the wedge of a frame $\theta^{\nu}$
\begin{equation}
(\ast{}R)_{\mu\nu}\wedge\theta^{\nu}
=\frac{1}{4}\epsilon_{\mu\nu\xi\kappa}R^{\xi\kappa}{}_{\lambda\sigma}\theta^{\lambda}\wedge\theta^{\sigma}
\wedge\theta^{\nu}=-G^{\rho}{}_{\mu}\eta_{\rho},
\end{equation}
where $G^{\rho}{}_{\mu}$ is the Einstein tensor and $\eta_{\rho}$
is the 3-form (i.e., $\eta_{\rho}=\ast\theta_{\rho}$).  Taking the
triple anti-symmetrization
\begin{eqnarray}
G^{\rho}{}_{\mu}\eta_{\rho}
&=&-\frac{1}{12}\left(\epsilon_{\mu\nu\xi\kappa}R^{\xi\kappa}{}_{\lambda\sigma}\theta^{\lambda}\wedge\theta^{\sigma}
\wedge\theta^{\nu}
+\epsilon_{\mu\lambda\xi\kappa}R^{\xi\kappa}{}_{\sigma\nu}\theta^{\sigma}\wedge\theta^{\nu}
\wedge\theta^{\lambda}
+\epsilon_{\mu\sigma\xi\kappa}R^{\xi\kappa}{}_{\nu\lambda}\theta^{\nu}\wedge\theta^{\lambda}
\wedge\theta^{\sigma}\right)\nonumber\\
&=&-\frac{1}{12}\left(\epsilon_{\mu\nu\xi\kappa}R^{\xi\kappa}{}_{\lambda\sigma}
+\epsilon_{\mu\lambda\xi\kappa}R^{\xi\kappa}{}_{\sigma\nu}
+\epsilon_{\mu\sigma\xi\kappa}R^{\xi\kappa}{}_{\nu\lambda}\right)
\theta^{\lambda}\wedge\theta^{\sigma}
\wedge\theta^{\nu}.\label{21eMay2010}
\end{eqnarray}
As the Einstein tensor vanishes in vacuum, the result follows.

\section{Tensorial proof for the completely symmetric of the Bel-Robinson tensor}
Here we present a detailed proof of the complete symmetry of the
Bel-Robinson tensor using the basic tensorial method. Consider
\begin{eqnarray}
T_{\alpha\beta\mu\nu}&:=&R_{\alpha\lambda\mu\sigma}R_{\beta}{}^{\lambda}{}_{\nu}{}^{\sigma}
+\ast{}R_{\alpha\lambda\mu\sigma}\ast{}R_{\beta}{}^{\lambda}{}_{\nu}{}^{\sigma}\nonumber\\
&=&R_{\alpha\lambda\mu\sigma}R_{\beta}{}^{\lambda}{}_{\nu}{}^{\sigma}
+\frac{1}{2}\epsilon_{\alpha\lambda\xi\kappa}R^{\xi\kappa}{}_{\mu\sigma}
\frac{1}{2}g_{\tau\beta}g_{\gamma\nu}\epsilon^{\tau\lambda\rho\pi}R_{\rho\pi}{}^{\gamma\sigma}
\nonumber\\
&=&R_{\alpha\lambda\mu\sigma}R_{\beta}{}^{\lambda}{}_{\nu}{}^{\sigma}
+\frac{1}{2}\epsilon_{\alpha\lambda\xi\kappa}R^{\xi\kappa}{}_{\mu\sigma}
\frac{1}{2}g_{\tau\beta}g_{\gamma\nu}\left(-\epsilon^{\tau\gamma\rho\pi}R_{\rho\pi}{}^{\sigma\lambda}
-\epsilon^{\tau\sigma\rho\pi}R_{\rho\pi}{}^{\lambda\gamma}\right)
\nonumber\\
&=&R_{\alpha\lambda\mu\sigma}R_{\beta}{}^{\lambda}{}_{\nu}{}^{\sigma}
+\frac{1}{2}\epsilon_{\alpha\lambda\xi\kappa}R^{\xi\kappa}{}_{\mu\sigma}
\frac{1}{2}g_{\tau\beta}g_{\gamma\nu}\left(-\epsilon^{\sigma\lambda\rho\pi}R_{\rho\pi}{}^{\tau\gamma}
+\epsilon^{\tau\sigma\rho\pi}R_{\rho\pi}{}^{\gamma\lambda}\right)
\nonumber\\
&=&R_{\alpha\lambda\mu\sigma}R_{\beta}{}^{\lambda}{}_{\nu}{}^{\sigma}
+\frac{1}{4}\delta^{\sigma\lambda\rho\pi}_{\alpha\lambda\xi\kappa}
g_{\tau\beta}g_{\gamma\nu}R^{\xi\kappa}{}_{\mu\sigma}R_{\rho\pi}{}^{\tau\gamma}
+\ast{}R_{\alpha\lambda\mu\sigma}\ast{}R_{\nu}{}^{\lambda}{}_{\beta}{}^{\sigma}
\nonumber\\
&=&\left(\frac{1}{2}R_{\alpha\mu\lambda\sigma}R_{\beta\nu}{}^{\lambda\sigma}
+R_{\alpha\lambda\mu\sigma}R_{\nu}{}^{\lambda}{}_{\beta}{}^{\sigma}\right)
-\frac{1}{2}R_{\alpha\mu\lambda\sigma}R_{\beta\nu}{}^{\lambda\sigma}
+\ast{}R_{\alpha\lambda\mu\sigma}\ast{}R_{\nu}{}^{\lambda}{}_{\beta}{}^{\sigma}
\nonumber\\
&=&R_{\alpha\lambda\mu\sigma}R_{\nu}{}^{\lambda}{}_{\beta}{}^{\sigma}
+\ast{}R_{\alpha\lambda\mu\sigma}\ast{}R_{\nu}{}^{\lambda}{}_{\beta}{}^{\sigma}
\nonumber\\
&=:&T_{\alpha\nu\mu\beta}.\label{10bJune2010}
\end{eqnarray}
From (\ref{10cJune2010}) and (\ref{10bJune2010}), the completely
symmetric property easily follows.  After we found this method we
learned that which is similar to the ones used in Senovilla paper
in 2000~\cite{Senovilla} (see Proposition 6.3).

However, can now we propose another easier proof the completely
symmetric of the Bel-Robinson tensor?  It is possible.  The idea
is simply making use of the property indicated
in~(\ref{29aJune2010}), the formal dual of the first Bianchi
identity (which is only valid in vacuum).  The basic idea is that
we can treat the formal dual of the Riemann curvature tensor as
the usual tensor manipulation.  For example, we know
$R_{\alpha\beta\mu\nu}=R_{\mu\nu\alpha\beta}$ in general and this
can immediately to treat the relation in vacuum
$\ast{}R_{\alpha\beta\mu\nu}=\ast{}R_{\mu\nu\alpha\beta}$, this is
another representation of the standard left dual and right dual
denoted in (\ref{30bJune2010}).  Although the idea and the
associated result is simple, it turns out that it is very useful
and practical. Here we give the detail derivation of the
completely symmetric of the Bel-Robinson tensor as follows
\begin{eqnarray}
T_{\alpha\mu\beta\nu}-T_{\alpha\nu\beta\mu}
&=&R_{\alpha\lambda\beta\sigma}R_{\mu}{}^{\lambda}{}_{\nu}{}^{\sigma}
-R_{\alpha\lambda\beta\sigma}R_{\nu}{}^{\lambda}{}_{\mu}{}^{\sigma}
+\ast{}R_{\alpha\lambda\beta\sigma}\ast{}R_{\mu}{}^{\lambda}{}_{\nu}{}^{\sigma}
-\ast{}R_{\alpha\lambda\beta\sigma}\ast{}R_{\nu}{}^{\lambda}{}_{\mu}{}^{\sigma}
\nonumber\\
&=&\frac{1}{2}R_{\alpha\beta\lambda\sigma}R_{\mu\nu}{}^{\lambda\sigma}
+\frac{1}{2}\ast{}R_{\alpha\beta\lambda\sigma}\ast{}R_{\mu\nu}{}^{\lambda\sigma}
\nonumber\\
&=&0,
\end{eqnarray}
using (\ref{30aJune2010}).  Consider the first two terms,
employing the Bianchi identity $B_{\alpha[\beta\mu\nu]}=0$
\begin{equation}
R_{\alpha\lambda\beta\sigma}(R_{\mu}{}^{\lambda}{}_{\nu}{}{}^{\sigma}-R_{\nu}{}^{\lambda}{}_{\mu}{}^{\sigma})
=R_{\alpha\lambda\beta\sigma}R_{\mu\nu}{}^{\lambda\sigma}
=\frac{1}{2}R_{\alpha\beta\lambda\sigma}R_{\mu\nu}{}^{\lambda\sigma}.
\end{equation}
And in the same formal way $\ast{}R_{\alpha\lambda\beta\sigma}
(\ast{}R_{\mu}{}^{\lambda}{}_{\nu}{}{}^{\sigma}-\ast{}R_{\nu}{}^{\lambda}{}_{\mu}{}^{\sigma})$
becomes
$\ast{}R_{\alpha\lambda\beta\sigma}\ast{}R_{\mu\nu}{}^{\lambda\sigma}$
which is
$\frac{1}{2}\ast{}R_{\alpha\beta\lambda\sigma}\ast{}R_{\mu\nu}{}^{\lambda\sigma}$

Moreover, using Lanczos identity, there is a well known equation
in empty space~\cite{Lovelock}
\begin{equation}
R_{\alpha\lambda\sigma\tau}R_{\beta}{}^{\lambda\sigma\tau}
\equiv\frac{1}{4}g_{\alpha\beta}R_{\rho\lambda\sigma\tau}R^{\rho\lambda\sigma\tau}.\label{10fJune2010}
\end{equation}
One can verify this identity in vacuum by employing an orthonormal
frame.  We first define
\begin{equation}
E_{ab}:=R_{0a0b}, \quad H_{ab}:=\ast{}R_{0a0b},
\end{equation}
where $E_{ab}$ and $H_{ab}$ are the electric and magnetic parts of
the Weyl tensor in vacuum.  Indeed we have verified this identity
recently \cite{So2010arXiv}.  It is a simple straightforward but
tedious calculation.  However, instead of making use the Lanczos
identity, we can reproduce this result using a simple tensorial
method similar to the above. From the completely symmetric
property, simply consider the two cases
\begin{eqnarray}
T_{\alpha\beta\mu\nu}&:=&R_{\alpha\lambda\mu\sigma}R_{\beta}{}^{\lambda}{}_{\nu}{}^{\sigma}
+R_{\alpha\lambda\nu\sigma}R_{\beta}{}^{\lambda}{}_{\mu}{}^{\sigma}
-\frac{1}{2}g_{\alpha\beta}R_{\mu\lambda\sigma\tau}R_{\nu}{}^{\lambda\sigma\tau},\label{10dJune2010}\\
T_{\mu\nu\alpha\beta}&:=&R_{\alpha\lambda\mu\sigma}R_{\beta}{}^{\lambda}{}_{\nu}{}^{\sigma}
+R_{\alpha\lambda\nu\sigma}R_{\beta}{}^{\lambda}{}_{\mu}{}^{\sigma}
-\frac{1}{2}g_{\mu\nu}R_{\alpha\lambda\sigma\tau}R_{\beta}{}^{\lambda\sigma\tau}.\label{10eJune2010}
\end{eqnarray}
We know that these two equations are equivalent in vacuum,
consider the last terms of (\ref{10dJune2010}) and
(\ref{10eJune2010})
\begin{equation}
g_{\alpha\beta}R_{\mu\lambda\sigma\tau}R_{\nu}{}^{\lambda\sigma\tau}
\equiv{}g_{\mu\nu}R_{\alpha\lambda\sigma\tau}R_{\beta}{}^{\lambda\sigma\tau}.
\end{equation}
Either taking the trace on $g_{\alpha\beta}$ or $g_{\mu\nu}$, the
result appears as shown in (\ref{10fJune2010}).

\section{Conclusion}
The Bel-Robinson tensor possesses many nice properties such as
completely symmetric.  Penrose used spinors to prove that indeed
it is true.  We have a proof to show that $T_{\alpha\beta\mu\nu}$
has this symmetric property.  Soon after we have this result we
learned that Senovilla in 2000 has the result using a similar
method as we did.  However, keep using the tensorial method and
making use the formal dual of left and right, we propose another
easier proof that $T_{\alpha\beta\mu\nu}$ is indeed completely
symmetric.

Here we provide a basic and straightforward tensorial method to
verify that the Bel-Robinson tensor is really completely
symmetric.  One may ask why we prefer a tensorial method to prove
something that it was well known.  In particular, why not keep
using spinors.  The reason is that although spinors are very power
and elegant, it may be worthwhile using a basic and simple
tensorial way to understand this symmetric property especially for
someone who is not familiar with spinors.  In fact, people usually
learning general relativity starts with the tensorial and then
study spinors afterwards. This is the reason we presented this
simple verification for the symmetric property of the Bel-Robinson
tensor.

Moreover, during the proof of the completely symmetric property of
the Bel-Robinson tensor, we have discovered something extra, which
is the well known relation in vacuum
$R_{\alpha\lambda\sigma\tau}R_{\beta}{}^{\lambda\sigma\tau}
=\frac{1}{4}g_{\alpha\beta}R_{\rho\lambda\sigma\tau}R^{\rho\lambda\sigma\tau}$.
It is amazing that we can reproduce this result only using such a
simple tensorial method.  Of course, one can simply use
orthonormal frames to verify this identity, but it is a testing
method fundamentally.  In other words, it is not a deduction.
However, making use of the known symmetric property of the
Bel-Robinson tensor, we have recovered the one-quarter identity.
This is a simple and nice proof.

After using the tensorial method with some successful results, one
may wonder whether the basic tensorial method is very useful so
that we do not need any other method.  In particular, spinors. The
answer seems negative.  This is because the tensorial method has
its own limitations.  There are still some things that using the
tensorial method are not easy to proof.  Then we have to use
another method. This may be the reason why people invented spinors
and used then for a long time.

\section*{Acknowledgment}
This work was supported by NSC 98-2811-M-008-078.

\end{document}